\newcommand{\degrees}{^\circ}
\documentclass[letterpaper,12pt]{article}   
\usepackage{osajnl2} 
\usepackage[draft]{hyperref} 
\usepackage{amssymb}
\usepackage{amsmath}

\begin{document}

\title{A compact and robust method for full Stokes spectropolarimetry}


\author{William Sparks,$^{1,*}$ Thomas A. Germer,$^2$  John MacKenty$^1$ and Frans Snik$^{3}$}
\address{$^1$Space Telescope Science Institute, 3700 San Martin Drive, Baltimore, MD 21218, USA}
\address{$^2$National Institute of Standards and Technology, 100 Bureau Drive, Gaithersburg, MD 20899, USA}
\address{$^3$Sterrewacht Leiden, Universiteit Leiden, Oortgebouw 570, Niels Bohrweg 2, 2333 CA Leiden, The Netherlands}
\address{$^*$Corresponding author: sparks@stsci.edu}

\begin{abstract}
We present an approach to spectropolarimetry which requires neither
moving parts nor time dependent modulation, and which offers the
prospect of achieving high sensitivity.
The technique applies  equally well, in principle, in the optical, UV or IR.
The concept, which is one of
those generically known as channeled polarimetry, is to encode the
polarization information at each wavelength along the spatial
dimension of a 2D data array using static, robust optical
components. A single two-dimensional data frame contains the full
polarization information and can be configured to measure either two
or all of the Stokes polarization parameters. By acquiring full
polarimetric information in a single observation, we simplify
polarimetry of transient sources and in situations where
the instrument and target are in relative motion. The robustness and
simplicity of the approach, coupled to its potential for high
sensitivity, and applicability over a wide wavelength range,
 is likely to prove useful for applications in challenging
environments such as space.
\end{abstract}

\ocis{120.5410, 260.5430.}

\maketitle 

\section{Introduction}
The polarization of light provides a versatile suite of remote sensing
diagnostics.  In astronomy, polarization is used to study the Sun and
Solar System, stars, dust, supernova remnants, and high-energy
extragalactic astrophysics\cite{SnikKeller2012}.  The astrophysical
mechanisms by which polarized light is produced range from scattering
phenomena to the interaction between high energy charged particles, and
magnetized plasmas.  Beyond astronomy, polarization is used in remote
sensing, medical diagnostics, defense, biophysics, microscopy, and
fundamental experimental physics, e.g. \cite{Goldstein2011}.

Accurate, precision polarimetric methods usually require rapidly
modulating, often fragile, parts and are inherently monochromatic,
e.g. photoelastic modulators (PEMs), ferroelectric liquid crystals or
liquid crystal variable retarders (LCVRs) in tandem with phase locked
photomultipliers, or synchronized charge shuffling on a charge-coupled device (CCD) detector
for area detection\cite{SnikKeller2012}.  Lower accuracy techniques
typically require sequential measurements of the target using
rotating waveplates and polarization analyzers.  Here we describe a
method to encode polarimetric information over a wide spectrum in
a single data frame, using static optics.  This approach alleviates
errors introduced by the need to match sequentially acquired data,
and eliminates the need for fragile or rapid modulation, yet may be able
to accomplish high accuracy, precision measurements.  The methods, of
course, have their own implicit sensitivities and concerns, as we
discuss below.

A particular interest of the authors, which serves as a useful
illustrative example, is the use of precision
circular polarization spectroscopy as a remote sensing
biosignature and a potentially valuable tool in searches for
biological processes elsewhere in the Universe.  
The circular polarization spectrum is sensitive to the presence of
molecular homochirality, a strong biosignature, through the combined
optical activity and homochirality of biological
molecules\cite{Sparks2009pnas, Sparks2009jqsrt}. Biologically-induced
degrees of circular polarization have been found in the range $10^{-2}$ to $10^{-4}$
for a variety of photosynthetic samples, with an important correlation
between the intensity spectrum and polarization
spectrum\cite{Sparks2009pnas}. Hence, precision full Stokes
polarimetry and wide spectral coverage are required.  Furthermore, the target scene
and instrumentation may be in rapid relative motion, compounding the
difficulties of acquiring the data using traditional polarimetric
techniques. A large number of photons must be accumulated in a
short period of time. The techniques presented in this paper may
provide a means to make this type of polarization measurement, in
addition to providing a robust method for acquiring less precise
spectropolarimetry in a straightforward fashion.
Furthermore, the approach is applicable across a wide wavelength range, and,
as well as in the visible, can work equally well in the ultraviolet, where for example chiral electronic signatures are
generally strongest, to the infrared, where polarimetry goes hand in hand with probes into
the geometry and physical characteristics of dusty regions of the universe.

A variety of similar concepts are available under the generic title of
``channeled polarimetry''\cite{Goldstein2011}. These typically fall
into two classes: channeled imaging polarimetry (CIP) and channeled
spectropolarimetry (CS), following terminology of
\cite{Goldstein2011}. To simplify, the CS methods typically encode the
polarization information as an amplitude modulation directly on the
spectrum, derived from a polarization optic whose retardance is a
function of wavelength. As an example, the spectral modulation
principle for linear spectropolarimetry\cite{Snik2009} can reach
a precision of at least $2\times 10^{-4}$\cite{vanHarten2011}. The CIP
methods, by contrast, use a polarization optic whose retardance is
spatially varying, so that the polarization information is encoded as a set
of spatial fringes onto an image\cite{Oka2003}. These two approaches, 
as well as a number of technical issues that arise in each case,
are described in some detail in \cite{Goldstein2011}. Previous authors
have used multiorder retarders, birefringent wedges, pairs of
birefringent wedges, and Savart plates individually or in combination
for these two applications \cite{Snik 2009, Oka2003, Serkowski1972,
  Nordseick1974, Oka1999, Oka2006, Mujat2004, Wakayama2008,
  Howard2008, Kudenov2008, Snik2010}. Typically, the polarization
information is extracted from the data using Fourier methods. Another approach to
single-shot imaging polarimetry and spectropolarimetry is the wedged
double Wollaston device, which yields multiple images on a detector
with polarization axes at different angles and allows retrieval of the
Stokes parameters through combinations of the images
\cite{Geyer1996,Oliva1997,Pernechele2003}.

The approach explored in the current paper is to
disperse the spectral and polarimetric information along two orthogonal directions,
a ``spectral'' dimension for the spectroscopy and a ``spatial'' dimension for the polarimetry. 
The amplitude modulation of the
encoding of the polarization information is independent of the
choice of spectral resolution. The two aspects of the measurement,
the spectroscopy and the polarimetry, may be optimized
independently. The complete spectropolarimetric information is encoded
on a single data frame, and may be derived using straightforward
analytical techniques.

Poisson photon counting statistics play a critical role in
astronomical polarimetry. To measure a polarization degree of
$10^{-n}$, it is necessary to collect (at least) $10^{2n}$
photons. For example, to measure $p\approx 10^{-4}$, it is necessary
to accumulate $10^8$ photons. A typical astronomical CCD has
a well-depth $\sim 10^5$ electrons per pixel, requiring $10^3$
pixel readouts. If the data are needed in, say, 1~s in one pixel, this multi-readout approach
becomes prohibitive.  A solution is to spread the illumination across
many pixels, as is done for high signal-to-noise-ratio photometry with
the Hubble Space Telescope \cite{Brown2001}. Making a virtue of necessity, if we
use optics which spread the light of a spectrum perpendicular to the
spectrum, then we can exploit the width of the broadened spectrum to
encode the polarimetric data.

Sections 2 and 4 discuss a variety of configurations that accomplish
this goal.
Sec.~2 starts with linear polarization (equivalently any two of
the Stokes polarization parameters), followed by a
discussion of our analysis methods in Sec.~3. Configurations that enable full
Stokes spectropolarimetry are presented in Sec.~4. Sec.~5 describes practical
implementation, sensitivities, and an approach based on calibration.
Sec.~6 provides an example application. Finally, we make some conclusions in Sec.~7.

The different embodiments of the underlying approach described in Secs.~2 and 4
highlight different aspects of the method. In the end, we anticipate that the most useful
realizations of the concept will be the double wedge for linear polarimetry, Sec.~2B, and
the double-double wedge for full Stokes spectropolarimetry, Sec.~4.B.3.
The other subsections introduce new ideas incrementally, while these two sections
capture the final products for the two types of polarimetry.

We use the conventional Stokes vector formalism to quantify the
polarization of light with $S\equiv (I,Q,U,V)$ where $I$ is the total
intensity; $Q, U$ decribe the linear polarization and $V$ the circular
polarization. The normalized Stokes parameters $(q,u,v)=(Q,U,V)/I$
represent the fractional polarization state. The degree of
polarization is given by $p=\sqrt{q^2+u^2+v^2}$ and the direction of
linear polarization by $\psi = {1\over 2}\tan^{-1}(u/q)$.

\section{Concept for Linear Polarimetry}

We envisage a spectrum of light broadened in a direction orthogonal to
the dispersion direction and sensed using a two dimensional area array
such as a CCD. This broadening can be spread along a segment of a conventional long slit
spectrograph, for example, with the length of the entrance slit
providing the spatial dimension in the detected two dimensional
spectrum. To introduce amplitude modulation along the slit ($x$
direction), we introduce a retardance gradient $\phi(x)$ along $x$
using a birefringent wedge (or wedges) followed by
a polarization analyzer, such as a dichroic polarizer or polarizing
prism (see Figs.~1 and 2).  It would be possible to carry out the
polarization analysis immediately in front of the detector array, as
in \cite{Oka2003}. However, performing the
polarization analysis as early as possible in the optical path yields
better robustness against polarization introduced by the
instrumentation optics. Furthermore, the polarization optics can be
more compact and easier to characterize, since light from all wavelengths
is analyzed using the same optical elements. Hence,
we prefer to insert the polarization optics
immediately adjacent to the spectrograph's entrance slit, and allow a long-slit spatial segment to
project through the instrument to become the detector array's spatial dimension.

\subsection{Single birefringent wedge}

To lay the groundwork, we initially consider just a single
birefringent wedge. The wedge thickness gradient is oriented along the
slit, while its fast axis is oriented $45^{\circ}$ with respect to the
slit. The analyzer's transmission axis is parallel to the slit, though it could
alternatively be orthogonal to it.
If we define the Stokes $Q$ direction as also being
parallel to the slit, then we can consider a uniformly illuminated slit with the
beam entering the slit, orthogonal to the slit plane. If the incoming
light is polarized with its electric field along the slit ($q=1$), at the
hypothetical tip of the wedge where the retardance is zero, the
polarized light passes through the retarder and analyzer without
hinderance. Moving along the slit, the retardance increases to the
point where it becomes quarter-wave, and the light is converted to
circularly polarized light after the retarder, and half of the light
transmits through the analyzer. As $x$, and  the retardance,
increase together, it reaches the point where there is half wave
retardance. At that point the polarization is rotated $90^\circ$ after
the retarder, and none of the light transmits through the analyzer.
At the same distance further along the slit, the retardance is
full-wave, the light is completely transmitted, and the cycle is
complete. Note that for typical birefringent materials, the spatial
distance $x$ corresponding to one wavelength of retardance will depend
on the wavelength. In the absence of dispersion, the spatial modulation
frequency is $\propto 1/\lambda$.

Circularly polarized light ($v=1$) is half transmitted at zero
retardance (circularly polarized light passing through the
analyzer). When the retardance reaches quarter-wave, the light becomes
linearly polarized along the slit direction and all of the light
passes the analyzer. When the retardance is half-wave, the sign of the
circular polarization is flipped ($v=-1$), and again, half of the
light is transmitted through the analyzer.  Hence, the modulation for
$v=1$ is similar to the modulation due to $q=1$, but out of phase by a
distance corresponding to one quarter wave of retardance.  Light
polarized linearly at $45^{\circ}$ ($u=1$), i.e., along the retarder
fast axis, is unaffected by the variable retardance along the slit.
However, if we precede the birefringent wedge by an achromatic (or
superachromatic) quarter wave retarder, with fast axis {\it along} the
slit, then the circular polarization parameter $V$ is interchanged
with $U$ \cite{Snik2009}. Now $V$ is unaffected by the variable
retardance and causes no spatial modulation, while $U$ causes spatial
amplitude modulation, a quarter wave out of phase from $Q$. If the
input Stokes vector is $(I,Q,U,V)$, the output intensity
$I_{\parallel}$ in the spatial direction is
\begin{equation}
I_{\parallel} = 0.5 (I + Q \cos\phi + U \sin\phi).
\end{equation}
The retardance $\phi$ maps onto the distance $x$ along the slit
according to $\phi = 2\pi(x/X)$, where the distance corresponding to
a single wave of retardance change is $X = \lambda/(|n_o -
n_e|\tan\xi)\}$, $\xi$ is the wedge angle, and $n_o$ and $n_e$ are the refractive
indices for the $o$ and $e$-beams, respectively.  If the circular polarization $V$
is desired rather than $U$, then the quarter wave retarder can be
omitted.

If a beam splitting polarizing prism (e.g. a Wollaston prism) is used
as the polarization analyzer, then two versions of the spectra are
obtained on the detector. The intensity $I_{\perp} $ of the
orthogonally polarized beam is
\begin{equation}
I_{\perp} = 0.5 (I - Q \cos\phi - U \sin\phi).
\end{equation}
The difference of Eqs.~(1) and (2), divided by their sum, assuming any
transmission differences have been removed, gives
\begin{equation}
I'=(I_{\parallel}-I_{\perp} )/(I_{\parallel}+I_{\perp} )= q \cos\phi + u \sin\phi.
\end{equation}
Table~1 summarizes this and other algebraic expressions for the
spatial modulation in subsequent configurations discussed below.  An
alternative way to express Eqs.~(1)--(3) is
\begin{eqnarray}
I_{\parallel} &= &0.5I(1 + p\cos(\phi - 2\psi))\nonumber\\
I_{\perp}  &= &0.5I(1 - p\cos(\phi - 2\psi))\nonumber\\
I' &=& p\cos(\phi - 2\psi),
\end{eqnarray}
where the position angle $\psi$ of linear polarization is given by
$\psi = {1\over 2}\tan^{-1}(U/Q)$. From Eq.~(4) it is apparent that
the spatially modulated profile has an amplitude of modulation equal
to the degree of polarization, and a (spatial) phase zero point that
reveals the angle of polarization.

It is important in using Eqs.~(1) and (2) that there not be any
signficant intensity variations along the slit on length scales of
order $X$. However, in the ``dual beam'' version, Eq.~(3), the total
intensity $I$ along the slit has been eliminated \cite{Snik2009}. This
potentially offers a means to retain some spatial resolution along the
slit. For example, the image of a star can have very large intensity
changes along a spectrograph slit, though its polarization is
unchanged. Provided the extent of the image is sufficient to encode
the sine and cosine terms in Eq.~(3), we may derive its polarization
even in the presence of quite strong intensity changes. In Eqs.~(1)
and (2), intrinsic intensity changes would be mixed with amplitude
modulation produced by polarization. Hence, care needs to be taken
in matching the spatial extent of the instrumental point spread function
to the projected scale of the retardance variation. Use of Eq.~(3) is
more robust against this constraint.

In practice, wedge components available off the shelf are relatively
thick. This introduces a multiorder retarder effect, exploited in the
SPEX concept \cite{Snik 2009}. A thick birefringent material, followed
by analysis optics, such as those employed here, for a single location
on the slit, yields spectral modulation,  used to measure the polarization in \cite{Snik2009}.
Hence,
using a single wedge, the resulting fringes from polarized light have
a relatively pronounced ``slope'' because the retardance, implicitly,
is varying as a function of both wavelength and spatial direction.
From above, the complete expression for $\phi$ is $\phi =
2\pi x(n_o - n_e)\tan\xi / \lambda$.
Therefore a constant phase $\phi$, which
defines the appearance of the fringes on the detector, occurs for $x =
\left[ \phi/(2\pi (n_e-n_o)\tan\xi)\right]\lambda$.  The thickness of
the wedge on the narrow edge effectively adds a constant to $\phi$ and
hence increases the slope of $x$ vs. $\lambda$.

In principle, analysis methods applied to the orthogonal dimension,
discussed in Sec.~4, can remove this spectral modulation. The slope
also translates to a constraint on the spectral resolution, as it must be sufficient to separate the fringes.

\subsection{Double wedge}

An alternative strategy to alleviate the spectral resolution
constraint, and at the same time simplify and render the analysis more
robust, is to compound the first wedge with an identical second wedge
reversed in direction, with its fast axis orthogonal to the fast axis
of the first wedge. The resulting optic, also used in Babinet-Soleil
compensators, is convenient to work with (it is now a rectangular
cuboid, see Fig.~3), and the reversed signs of birefringence conspire
to yield a zero retardance at the point where the two retardances are
equal, expected to be close to the center of the optic. This is the
configuration discussed by \cite{Goldstein2011, Oka2003} in the
context of imaging polarimetry. In imaging polarimetry, spectral
modulation across the bandpass would be highly problematic, and this
configuration goes some way towards eliminating
it. \cite{Goldstein2011} discusses other ways to mitigate
birefringence dispersion. When applied to spectropolarimetry, as here,
the effect is to remove the slope from the fringes and cause them to
be approximately parallel in the spectral dimension, with spacing
convergence to the blue, as a given amount of retardance corresponds to
a greater number of wavelengths.

Using Mueller matrix algebra, it can be shown that the amplitude
modulated profiles for this configuration are, for the single beam,
with a quarter wave plate used, as above, to impose a sensitivity to Stokes~$U$
rather than Stokes~$V$:
\begin{equation}
I _{\parallel}= 0.5 (I + Q \cos 2\phi + U \sin\ 2\phi)
\end{equation}
and for the dual beam
\begin{equation}
I'=q \cos 2\phi + u \sin 2\phi.
\end{equation}
The essential characteristics of the compound wedge profiles are the
same as the single wedge, though the spatial frequency is doubled and
the multiorder retarder effect with wavelength is removed making the
fringes essentially parallel to the dispersion direction. The double
wedge also yields a quasi-zero order retarder, which has much
smaller temperature dependence than a single wedge.

Fig.~4 shows simulations of data frames obtained with
100\% polarized light for each of the configurations discussed here,
for $Q$, $U$ and $V$ polarized light separately.  To illustrate the
concepts of this section, Fig.~5 shows actual long-slit
spectra obtained using one and two quartz birefringent wedges with a
$3^{\circ}$ wedge angle, installed in a simple slit
spectrograph. We inserted the quartz wedges immediately after the
entrance slit of the spectrograph, together with a quarter wave plate
and analyzer as discussed. Fig.~5 shows fringes obtained when
illuminated by linearly polarized light for both single wedge and
double wedge. These correspond to configurations $w$ and $ww^\prime$
defined in Table~1, though the wedges are not compounded in
this test. The fringes are clearly visible when polarized light enters
the spectrograph, and are not visible when unpolarized light is used. A more
formal laboratory validation follows below in Sec.~6.

\section{Data analysis Methods}

In equations (1)--(3), (5), and (6), the Stokes parameters are
coefficients of orthogonal trigonometric functions. While
Fourier methods could be used to retrieve these coefficients, we
prefer a linear least squares solution. This enables us to generate
formal error estimates, in addition to providing the Stokes
coefficients in a straightforward fashion.  With a least squares
approach, one can also take the dispersion of birefringence into
account in a more straightforward way than with Fourier methods.  The
general methods are described in the Appendix. If the spatial profile
at a fixed wavelength is given by
\begin{equation}
I_{\hbox{obs}} = I i_c + Q q_c + U u_c + V v_c
\end{equation}
for a single beam, or
\begin{equation}
I' = (I _{\parallel}-I _{\perp})/(I _{\parallel}+I _{\perp})=q\cdot q_c + u\cdot u_c + v\cdot v_c
\end{equation}
for the dual beam formalism, then the terms $i_c$, $q_c$, $u_c$ and
$v_c$ are trigonometric functions whose coefficients are the Stokes
parameters we seek, i.e., Eqs.~(1)--(3), (5), (6), and others below. Since these
trigonometric functions take on a set of known values at each $x_i$
along the profile, standard methods can be applied to solve for their
coefficients given an observed intensity profile $I_{\hbox{obs}}(x_i)$
or $I'(x_i)$, as described in the Appendix . The solution depends on
the inverse of the curvature matrix, which is derived from products of
the functions $i_c$, $q_c$, $u_c$ and $v_c$. The solution is the
inverse of the curvature matrix, multiplied by a vector derived from
the observed profile and the same set of functions. The covariance
matrix is also the inverse of the curvature matrix, and uncertainties
on the Stokes parameters are taken as the square root of the diagonal
terms of that matrix.

Analytic expressions can be derived for the terms of the curvature
matrix and its inverse.  Replacing the summations by integrals over
complete periods, it may be seen that in a formal mathematical sense
the Stokes parameters are independent because the trigonometric
functions are orthogonal for most of the methods described. That is,
the integrals of their products become zero and the inverse of the
curvature matrix is diagonal.  Hence there is no formal covariance
between the terms. In an instrumental realization this should
translate to no formal cross-talk between the Stokes parameters. In
the cases where there are off-diagonal terms in the curvature matrix
inverse, we discuss this in the text.

The description so far is idealized, however.
In general,
summations are not over complete periods, and departures of the wedge
characteristics from those assumed, e.g. the exact value of the slope,
and sampling errors can all yield off-diagonal terms.  We discuss methods
to deal with such issues in Sec.~5. These terms can generally be
included in the analytic analysis, although the equations can become
complicated depending on which tolerances are explored. An example can be
found at the end of Sec.~4.B.3.

It will also be necessary to calibrate a given system.  The goal of
calibration can be either to determine the (hopefully) small
correction factors to the analytic formulae, or else a completely
empirical calibration approach can be adopted, as discussed in
Sec.~5.B.

To within an order of magnitude, we see in the analytic solutions, presented in
the Appendix, the expected $1/\sqrt{N_{tot}}$ dependence for the
sensitivity to which polarization can be measured, where $N_{tot}$ is
the number of detected photons.

\section{Concept for Full Stokes Polarimetry}

The use of precision circular spectropolarimetry as a
biosignature\cite{Sparks2009pnas} requires accurate, sensitive
measurement of the circular polarization spectrum. It is strongly
preferred that all Stokes parameters be measured in order to better
understand the physics involved and to guard against instrumental
cross-talk, where one Stokes parameter is measured incorrectly, through
instrumental effects, as another. Here, we describe two options to
acquire full Stokes polarimetry data. The first, briefly described, is
a brute force approach, where we simply place two versions of the
configurations described above, next to one another. The second
approach is to use an additional wedge or compound wedge with
different fast axes and wedge angles to fully encode all Stokes
parameters on a single data frame using a single optical bench.

\subsection{Two spectrograph  slits}

In Secs.~2A and 2B, Eqs.~(1)---(6), we showed how to encode two Stokes parameters
simultaneously.
By dispensing with the quarter wave retarder in those configurations, e.g. using a single wedge single beam, we have
\begin{eqnarray*}
I_{Q} = 0.5 (I + Q \cos\phi - V \sin\phi).
\end{eqnarray*}
If a second device is constructed with its wedge fast axis horizontal,
$0^{\circ}$, and analyzer at $45^{\circ}$ to the horizontal, then we
have
\begin{eqnarray*}
I_{U} = 0.5 (I + U \cos\phi + V \sin\phi).
\end{eqnarray*}
These two configurations can be processed indendently, and typically
$\vert V \vert \ll \vert Q\vert, \vert U \vert$. Following the
formalism of Eq.~(4), the influence of a component of circular
polarization $v$ is to shift the phases of the spatially modulated
profiles by amounts, in pixel space, of $\delta x =
X\tan^{-1}(v/q)/(2\pi)$ for the first device and $\delta x =
X\tan^{-1}(v/u)/(2\pi)$ for the second device, where $X$ is the
spatial period corresponding to one wavelength of retardance. For
small $v$, the spatial shift of the profile is therefore $\delta x
\approx Xv/2\pi q$ and $\delta x
\approx Xv/2\pi u$, respecively. For example, if the source is 3\% polarized,
$q=0.03$, if $v=10^{-3}$ (0.1\% circularly polarized), and if $X=10$~pixels, the spatial
shift is 0.05~pixels. The accuracy to which the
spatial shift can be determined depends in principle on the total
number of photons rather than the value of $X$, provided there are
sufficient points within $X$ to properly sample the profile (see
Appendix). Systematic instrumental effects are also likely to enter;
however, measurement to a precision of better than 0.01~pixels can be
achieved in precision astrometry, and these accuracies ought to be
feasible.

\subsection{Double wedges}

A concise solution to acquiring full Stokes spectropolarimetry may be
obtained by using a more complex group of birefringent wedges.
Our technique is analogous to the dual PEM polarimeters which encode the
different Stokes parameters using different carrier frequencies. If we
follow the first wedge or compound wedge by a second wedge that has twice the
thickness gradient, and a fast axis oriented $45^\circ$ to the first
one, then the resultant intensity profiles encode the full Stokes
information. The trigonometric functions involved are more complex
than the simple sine and cosines of above, but they are relatively
straightforward and still orthogonal.  This configuration has the advantage of
measuring the full Stokes parameters completely and simultaneously
for a source, without minor viewing perspective differences in the
double slit option (Sec.~4.A) and without needing to combine
measurements from two essentially independent polarimeters.

We describe three versions of this concept, though it is clear that a
variety of permutations and options are available under the umbrella
of this general approach. The first two versions, to illustrate the principles, are to
place a single wedge with twice the thickness gradient parallel and
antiparallel, respectively, after a single wedge oriented as for the linear
polarimetry application above. We conclude with a discussion of a double
compound wedge, in which the second pair
has twice the thickness gradient of the first. No quarter wave retarder is
required. An example of a different permutation would be to reverse
these two compound wedge pairs, which would result in a potentially
more convenient choice of analyzer angle following the wedges. We
defer discussion of such an option to a later paper and focus here on
providing a proof of concept and demonstration of the approach. The
naming convention used in the following subsection headings is
described in the footnote to Table~1, which summarizes the terms in the
equations for the amplitude modulation. Table~2 presents formulae for
the uncertainties on the normalized Stokes parameters derived using
the linear least squares methods of Sec.~3 and the Appendix.
Table~3 gives the error estimates for the unnormalized Stokes
parameters. Throughout, we assume standard   ($1\sigma$),
coverage
factor $k=1$, uncertainties.

\subsubsection{Two individual parallel wedges: $wW$}

In this version we assume the first wedge has a birefringence
gradient, described by $\phi(x)$, and fast axis at $45^{\circ}$ to the
horizontal, as above. We assume the second wedge has twice the
retardance $2\phi(x)$ at location $x$ and has fast axis aligned with
the slit, $0^{\circ}$. Additionally, it is necessary to allow the
analyzer angle to be set at angles other than zero degrees, in order
not to lose dependence to the Stokes $U$ parameter. We define the additional
variable $\theta$ as the angle of the transmission axis of the
analyzer with respect to the horizontal or slit direction. In the dual beam
configuration, the transmission axis of the second polarized beam is
$\theta + 90^\circ$. We retain the labelling $I_{\parallel}$ and
$I_{\perp}$ for these two beams, respectively.  In practice, there are
likely to be constant offsets and a gradient of the second wedge not
exactly twice that of the first wedge.  These terms can
generally be included in the analytic analysis of Sec.~3, but for
clarity in introducing the concepts, we set them aside for now.  It
can be shown that the amplitude modulated profiles for this
configuration are, for the single beam:
\begin{eqnarray}
I _{\parallel}= 0.5 (I +Q(\cos\phi\cos 2\theta+\sin\phi\sin 2\phi\sin 2\theta)+U\cos 2\phi\sin 2\theta\nonumber\\
    +V(\cos\phi\sin 2\phi\sin 2\theta-\sin\phi\cos 2\theta)),
\end{eqnarray}
\begin{eqnarray}
I _{\perp}= 0.5 (I -Q(\cos\phi\cos 2\theta+\sin\phi\sin 2\phi\sin 2\theta)-U\cos 2\phi\sin 2\theta\nonumber\\
    -V(\cos\phi\sin 2\phi\sin 2\theta-\sin\phi\cos 2\theta)),
\end{eqnarray}
and hence for the dual beam,
\begin{eqnarray}
I'= q(\cos\phi\cos 2\theta+\sin\phi\sin 2\phi\sin 2\theta)+u\cos 2\phi\sin 2\theta\nonumber\\
    +v(\cos\phi\sin 2\phi\sin 2\theta-\sin\phi\cos 2\theta).
\end{eqnarray}

We see that if $\theta=0^{\circ}$, then the coefficient of $U$ is zero
and hence we cannot derive the value of $U$. There is no particular
reason to have the analyzer at such an angle. For example, in Fig.~4 we use
$\theta=45^{\circ}$ for the double wedge configurations. However, the
derived variance for each of the Stokes parameters is a function of
$\theta$. Depending on the application, it may be helpful to
select $\theta$ to minimize the variance of the estimate for
Stokes $V$, since Stokes $V$ is usually orders of magnitude smaller
than the linear Stokes parameters. Below, Sec.~4.B.3 we show an
example.

\subsubsection{Two individual antiparallel wedges: $wW^\prime$}

If the second wedge is placed in the opposite direction to the first,
so as to minimize the geometric angles of the two wedges together, we have phase gradients
of $\phi(x)$ and $(\zeta-2\phi(x))$, respectively, for the two wedges, where $\zeta$ is a
constant, presumed unknown.
It can be shown that the amplitude modulated profiles for this configuration are, for the single beam:
\begin{eqnarray}
I _{\parallel}=  0.5  (I+Q(\cos\phi\cos 2\theta+\sin\phi\sin(\zeta -2\phi)\sin 2\theta)+U\cos(\zeta-2\phi)\sin 2\theta\nonumber\\
+V(\cos\phi\sin(\zeta -2\phi)\sin 2\theta-\sin\phi\cos 2\theta)),
\end{eqnarray}
\begin{eqnarray}
I _{\perp}= 0.5  (I-Q(\cos\phi\cos 2\theta+\sin\phi\sin(\zeta -2\phi)\sin 2\theta)-U\cos(\zeta-2\phi)\sin 2\theta\nonumber\\
-V(\cos\phi\sin(\zeta -2\phi)\sin 2\theta-\sin\phi\cos 2\theta)),
\end{eqnarray}
and for the dual beam
\begin{eqnarray}
I'= q\cos\phi\cos 2\theta+\sin\phi\sin(\zeta -2\phi)\sin 2\theta)-u\cos(\zeta-2\phi)\sin 2\theta\nonumber\\
-v(\cos\phi\sin(\zeta -2\phi)\sin 2\theta-\sin\phi\cos 2\theta).
\end{eqnarray}

In this case, where the second wedge is configured to run antiparallel
to the first one, there is an off-diagonal term in the inverse
curvature matrix ${\bf B}^{-1}$ (defined in the Appendix) containing
the term $(\sin\zeta\sin 4\theta)$. (The same term appears if the
wedges run parallel, but with a phase offset.)  The inverse of the
curvature matrix is the crucial mathematical entity in both solving
for the Stokes parameters and estimating their variances.  If the
off-diagonal terms are zero then the trigonometric functions are
orthogonal, and there are no formal dependencies of one Stokes
parameter on the others or covariances between them. If there are
off-diagonal terms, then there may be such dependencies
and covariances.

The off-diagonal term can be set to zero if the analyzer is
placed at an angle $\theta$ such that $\sin 4\theta$ is zero. Since
the $U$ coefficient also includes terms involving $\sin 2\theta$, which
we do not want to be zero, the desired analyzer angle to eliminate
cross-dependencies is $\theta=45^{\circ}$.
There may be a trade choice based on the optimization required,
between making the off-diagonal terms of the inverse curvature matrix
zero and minimizing the variance of a particular Stokes parameter,
which can occur at a different analyzer angle.

\subsubsection{Two compound wedge pairs: $ww^\prime WW^\prime$}

By analogy with the compound double wedge in the linear polarimetry
example, described in Sec.~2.B, it is possible to use two compound
double wedges to provide zero retardance in the center of the optic.
The first wedge pair has one gradient, as for the optic used in
Sec.~2.B above ($ww^\prime$ in the notation of Table~1). Each element
of the second wedge pair has double the thickness gradient. The first
has its fast axis along the slit, while the second has its fast axis
orthogonal to it, so that the retardances are $\xi-2\phi(x)$ and
$\xi+2\phi(x)$, respectively.  Again, the compound device with four
wedges is rectangular in shape.  It can be shown that the amplitude
modulated profiles for this configuration are, for the single beam,
\begin{eqnarray}
I _{\parallel}=  0.5(I+Q(\cos 2\phi\cos 2\theta+\sin 2\phi\sin 4\phi\sin 2\theta)+U\cos 4\phi\sin 2\theta\nonumber\\
+V(\cos 2\phi\sin 2\phi\sin 2\theta-\sin 2\phi\cos 2\theta)),
\end{eqnarray}
\begin{eqnarray}
I _{\perp}= 0.5(I-Q(\cos 2\phi\cos 2\theta+\sin 2\phi\sin 4\phi\sin 2\theta)-U\cos 4\phi\sin 2\theta\nonumber\\
-V(\cos 2\phi\sin 2\phi\sin 2\theta-\sin 2\phi\cos 2\theta)),
\end{eqnarray}
and for the dual beam
\begin{eqnarray}
I'= q(\cos 2\phi\cos 2\theta+\sin 2\phi\sin 4\phi\sin 2\theta)-u\cos 4\phi\sin 2\theta\nonumber\\
-v(\cos 2\phi\sin 4\phi\sin 2\theta-\sin 2\phi\cos 2\theta).
\end{eqnarray}

These functions are orthogonal to one another. Fig.~6 shows the derived
variance as a function of the analyzer angle, with analytic
formulae presented in Table~2. The minimum variance for Stokes $V$,
which typically has the smallest value of the Stokes parameters, is
given at analyzer angle with $\tan 4\theta = -2$, which implies
$\theta \approx 74.1^{\circ}$. The minimum variance for Stokes $Q$ is
at $\tan 4\theta = +2$, i.e.,  $\theta \approx 15.9^{\circ}$, and that
for $U$ is at $\theta=45^{\circ}$. The value of the minimum
variance is, for $q$ and $v$, $\sigma(q, v) \approx
1.24/\sqrt{N_{tot}}$, which represents the photon counting limit for
such a device and which is quite close to the canonical
$1/\sqrt{N_{tot}}$ value.

If the centers of the two compound wedges are misaligned by a spatial
distance $s$, then there is an additional term in the above equations,
which we characterize by a phase offset in the second wedge pair. That is,
their retardances are given by $\xi-2\phi(x)-a$ and
$\xi+2\phi(x)+a$, respectivel, where $a = 4\pi s/X$.  In this case, the amplitude
modulation expressions are
\begin{eqnarray}
I _{\parallel}=  0.5(I+Q(\cos 2\phi\cos 2\theta+\sin 2\phi\sin (2a+4\phi)\sin 2\theta)+U\cos (2a+4\phi)\sin 2\theta\nonumber\\
+V(\cos 2\phi\sin (2a+4\phi)\sin 2\theta-\sin 2\phi\cos 2\theta)),
\end{eqnarray}
\begin{eqnarray}
I _{\perp}= 0.5(I-Q(\cos 2\phi\cos 2\theta+\sin 2\phi\sin (2a+4\phi)\sin 2\theta)-U\cos (2a+4\phi)\sin 2\theta\nonumber\\
-V(\cos 2\phi\sin (2a+4\phi)\sin 2\theta-\sin 2\phi\cos 2\theta)),
\end{eqnarray}
and for the dual beam
\begin{eqnarray}
I'= q(\cos 2\phi\cos 2\theta+\sin 2\phi\sin (2a+4\phi)\sin 2\theta)-u\cos (2a+4\phi)\sin 2\theta\nonumber\\
-v(\cos 2\phi\sin (2a+4\phi)\sin 2\theta-\sin 2\phi\cos 2\theta).
\end{eqnarray}

Following through the least squares analysis, it can be seen that the
miscentering introduces an off-diagonal term in the inverse of the
curvature matrix, correlating the errors in $Q$ and $V$, and again
involving $\sin 4\theta$. Hence, this off-diagonal term can  be set
to zero by setting $\theta=45^{\circ}$. If this is not done, or the
tolerances do not allow it, then the presence of a covariance term in
and of itself does not bias the result. The random errors on $Q$
correlate with the errors on $V$. The actual values of $Q$ and $V$ do
not correlate, though, provided that the covariance term is known. In
principle the covariance term does contribute to the overall error
budget, although assuming independence yields a good description of
the actual observed variance in Monte Carlo simulations, shown in Fig.~7. In
the case where the term is present, but known, we still correctly
derive the Stokes parameters without cross-talk.

Cross-talk occurs if the covariance is not characterized correctly.
For the
example here of two miscentered wedge pairs, if we characterize the
miscentering by $a/X$, the ratio of the offcenter distance to the
spatial distance corresponding to one wavelength in the first wedge
pair, then the true value of Stokes $v$, say, is
$v_{true}=sy_1[B^{-1}]_{qv} + sy_4 [B^{-1}]_{vv}$, following the
notation of the Appendix. If we incorrectly derive $v_{est}=sy_4
[B^{-1}]_{vv}$, ignoring the cross-term, then
$v_{true}-v_{est}=sy_1[B^{-1}]_{qv}=q [B^{-1}]_{qv}/[B^{-1}]_{qq}$. A
formal tolerance analysis can be carried out, and to completely ignore
this term, while restricting the cross talk $<10^{-3}$, requires
$\theta$ to be within $\lesssim 0.1^\circ$ of $45^{\circ}$ if the
miscentering is such as to {\it maximize} the cross talk. If needed,
there are two options to relax this constraint within the context of
an analytical approach: (\romannumeral1) do not ignore the term! and
(\romannumeral2)~increase the spatial scale $X$ to ease the
requirement on $s/X$. Empirical calibration approaches are also
possible as discussed below.

\section{Practical Implementation, Requirements, Sensitivities and Calibration}

Wedge retarders, such as those discussed here may be custom
manufactured. However, testable quality versions are available off-the-shelf
under the guise of depolarizers or scramblers.  Quartz or calcite provide plausible
birefringent materials. The ordinary and extraordinary refractive
indices and birefringences are, for quartz, $n_o = 1.5384$, $n_e =
1.5473$, $n_e-n_o = 0.0089$, respectively, and for calcite $n_o
=1.647$, $n_e =1.480$, $n_e-n_o = -0.167$.  For example, if the
retarders have a wedge gradient of $3^{\circ}$, then at 500~nm the
retardance increases by one wavelength over a distance $X$ of $\approx
1.1$~mm for quartz and $0.06$~mm for calcite. If these scales are
projected 1:1 onto a detector with $5~\mu$m pixels, then these values correspond to
$X\approx 214$ and $11$ pixels for quartz and calcite,
respectively.

\subsection{An empirical calibration approach}

In a similar fashion to the analysis described in Sec.~4.B.3, the least
squares methods lend themselves to formal tolerance analyses, as
well as parameter estimation. However, a comprehensive analysis of all
plausibly relevant parameters is impractical and premature.  Instead,
we consider an alternative approach, which is purely empirical.  For a
very general set of optical component characteristics, the generic
versions of the amplitude modulated intensity profile, Eqs.~(7)
and (8), are valid, even if the exact forms of the functions $i_c$,
$q_c$, $u_c$ and $v_c$ are not known.  If we present the system in turn
with unpolarized light, and then 100\% polarized light oriented in
the $Q$ direction, the $U$ direction, and finally, with
100\% circularly polarized light, then the empirical response {\it
  gives} the functions $i_c$, $q_c$, $u_c$ and $v_c$. These
empirically derived functions can then be used to numerically derive the curvature
matrix and its inverse. As in Sec.~4.B.3, the
presence of covariance terms in the matrices by itself does not
invalidate the approach, because, in principle, with a high quality set
of calibration observations (which would appear much like the examples
shown in Fig.~4), they are implicitly known.  It is only when the
terms are not correctly accounted for, that problems may arise;
that is, if the calibration sources are not of high enough
quality.  It is also likely that the variance will be a function of
analyzer angle, as above. Hence empirical versions of Fig.~7 may be
useful to explore trade space.

We expect, in general, the empirical calibration method to be a very
important approach, potentially offering the best strategy for
deriving the Stokes parameters accurately. However it remains to be
seen to what degree the required tolerances, or calibration stability,
limit the performance of these devices in practice.  Our
intent is to develop additional laboratory experience to understand
these issues.

\subsection{Sensitivities}

Component birefringence will depend on temperature. This can be
mitigated by (\romannumeral1)~stabilizing the temperature,
(\romannumeral2)~continuous calibration, and
(\romannumeral3)~compounding carefully chosen materials.  To obtain an
idea of the order of magnitude of the temperature sensitivity, we use
the temperature dependent formula for the birefringence $B=n_e-n_o$ of
quartz, given by \cite{Beckers1967}, which yields $(dB/dT)/B\approx
1.2\times10^{-4}$ for the range $T=0$~$\degrees$C to
$25$~$\degrees$C. Since the wedge should keep its shape under
expansion or contraction, only the birefringence term matters and
imprints itself as an identical fractional change $\epsilon =
1.2\times10^{-4} dT$ on the spatial wavelength $X$. It can be shown
that the resulting fractional change in $q$ for the dual beam single
wedge example is one half the fractional change in $B$, for an input
beam consisting of purely $q$, and the spurious cross-talk $u'$
into the other Stokes parameter $u$ is $u'=\pi\epsilon q$.  Hence a
$1$~$\degrees$C temperature change would result in a spurious polarization
(cross talk) of $u'\approx 3\times 10^{-5}$ for a 10\% linearly
polarized source.

The presence of two refractive indices in a wedge-shaped optic will
cause a prismatic separation of the orthogonally polarized beams. The
magnitude of this effect will depend on the details of the optical
system designed. In our laboratory testing, this issue was not
significant.

If the beam incident on a single wedge has significant convergence,
then the retardance seen by light at different angles differs by approximately
$1/\cos \zeta$ where $\zeta$ is the angle to the normal. For this
retardance difference to be less than $\lambda/10$, say, the
$f$-number of the beam needs to be slower than $f/9$ for a 1~mm
thickness quartz and $f/40$ for 1~mm thickness calcite. The
convergence requirement scales as the inverse square root of the
thickness.

We carried out a number of Monte-Carlo simulations and found that the
retrieved solutions are close to the theoretical expressions given in
the tables provided that (\romannumeral1)~at least one full period is
sampled and (\romannumeral2)~there are sufficent sampling points
within the full cycle to properly sample the highest frequency
component.

In cases where the fringes have significant slope, as described above,
it will be necessary for the spectral resolution to be sufficient to
separate the fringes. If we require the retardance change
$\Delta\phi/(2\pi)<0.1$, this is
$(\partial\phi/\partial\lambda)\Delta\lambda<0.1$, where $\phi = 2\pi
L(n_e-n_o)/\lambda$, $L$ the thickness. Hence the required spectral
resolution is $R\equiv {\lambda\over \Delta\lambda} =
10L(n_e-n_o)/\lambda$. For 2~mm thick quartz this is $R>360$ and for
1~mm calcite, $R>3330$. For the compounded wedge optics which
straighten the fringes, this constraint is relaxed to the point of
being essentially irrelevant.

\section{Laboratory Validation}

We established an optical test bench to allow us to provide an
empirical proof of concept for the approach presented in this paper,
and to demonstrate that the device  functions
as a polarimeter.
The optical test bench is shown and illustrated in Fig.~8.

Light sources, either halogen continuum white
light, or a variety of line lamps for wavelength calibration, illuminated
an integrating sphere's entrance port.  The light
emerging from a separate port at right angles to the first is expected
to be unpolarized. However, to ensure an unpolarized source and to provide a
uniform location for our measurements, the emergent light was directed
to fall onto an opal diffusing screen.
Following the opal screen, and located
close to the screen, the light optionally encountered polarizing
elements ($Q$, $U$, or $V$) for calibration, samples to measure the
polarized transmission spectrum, or nothing, to provide an intensity
reference spectrum.

The spectrograph consisted of a slit, a collimator, a transmission grating,
and a camera. The slit was 125~$\mu$m by 1~cm and was located at one focus of an
$f/1.4$ 50~mm Nikon collimating lens.\footnote{Certain commercial
  equipment, instruments, or materials are identified in this paper in
  order to specify the experimental procedure adequately. Such
  identification is not intended to imply recommendation or
  endorsement by the Space Telescope Science Institute, the National
  Institute of Standards and Technology, or the Universiteit Leiden
  nor is it intended to imply that the materials or equipment
  identified are necessarily the best available for the purpose.} The 
transmission diffraction grating was ThorLabs Part \# GT50-06V with 600
grooves/mm. A second $f/1.4$ 50~mm Nikon lens imaged the spectrum onto 
a Quantum Scientific Imaging 683 8~Mpix cooled CCD Camera with $3326\times 2504$
pixels of size 5.4~micron.

For the polarization analyzer, we placed a Meadowlark precision linear
polarizer, DPM100VIS between the spectrograph slit and the collimating
lens, using a rotary mount to allow adjustment of the analyzer
angle. The extinction ratio across the visible spectrum for this
polarizer is $\sim 10^4$, and exceeds 100:1 from approximately 375~nm
to 725~nm.  The birefringent wedges were placed on a platform next to
the spectrograph slit between the slit and the source (i.e. before the
light enters the spectrograph).  The optional quarter wave retarder was
mounted, when used, between these wedges and the source and was an
achromatic Meadowlark quarter wave retarder AQM-100-545, which is
effective between 450~nm and 630~nm.  The birefringent wedges
themselves were a mix of customized and off-the-shelf quartz
scramblers from Karl-Lambrecht Corporation.  Off-the-shelf wedges had
a pitch angle of $\approx 3\degrees$, and the customized pieces used a
pitch angle of $\approx 6\degrees$.

The distance between the source and the spectrograph slit was
approximately 0.6~m, and the entire system was contained within a
series of light-tight baffles and boxes to eliminate stray light.
Locating the birefringent wedges externally to the spectrograph not
only allows us to ignore possible polarization due to the slit, but
also allowed easy access for switching configurations, and permitted
us to focus the spectrograph onto a single slit position.

The first exercise was to attempt to reproduce the appearance of the
theoretical data frames for the various configurations presented in
Fig.~4. Fig.~9 shows the results.  To do this, we used the white
halogen continuum source in combination with linearly and circularly
polarizing filters. For the first four configurations (first four
columns in Fig.~9), we used a set of 60~mm astronomical polarizing
filters that utilize HN38 Polaroid mounted in a magnesium fluoride
substrate to approximate 100\% linearly polarized light (first two
rows of the first four columns in Fig.~9). We used a cholesteric liquid
crystal technology (CLC) filter to approximate 100\% circularly polarized light (third row
of the first four columns in Fig.~9).  For the final configuration
(fifth column of Fig.~9), we used a polarization state generator that
utilized a precision linear polarizer in combination with a Fresnel
rhomb \cite{Boulbry2007} that is capable of producing close to
100\% polarized light anywhere on the Poincar\'e sphere.  We also obtained
data frames without any polarizing optics to provide a ``flat-field''
reference. The spectral scale was wavelength calibrated using argon and
mercury line lamps.  We defined the slit direction to correspond to
$Q$ and $45\degrees$ to the slit to correspond to $U$.

By comparing Figs.~4 and 9, it is apparent that the empirical data
reproduce the qualitative expectations extremely well. Differences in
detail can be attributed to different absolute thicknesses of the
wedges (in practice the 3$\degrees$ and 6$\degrees$ wedges had very
different thicknesses) relative to one another and to the theoretical
model, and to essentially random centering of the crossover points for
wedge pairs relative to the slit and to one another. Care was taken
with the sign convention and parity of the wedges to reproduce the
directions for fast axes and wedge gradients used in the models. The
empirical wavelength range shown corresponds to 550~nm to 700~nm, and the
slit height to 2~mm in Fig.~9.  We consider the data obtained in
this exercise to be fully consistent with the theoretical
expectations, given the practical uncertainties described.

The second exercise was to test the linear polarization mode described in
Sec.~2B. We placed a compound $3\degrees$ wedge pair with the fast
axes crossed, running corner to corner at $\pm 45\degrees$ to the
spectrograph slit, in front of the spectrograph slit. An achromatic
quarter wave retarder was placed upstream of the wedge pair, with its
fast axis oriented $0\degrees$ to the slit.  Calibration measurements were
taken using the standard suite of polarizing filters, and the
resulting data frames were used to derive empirical ``coefficient''
frames for use in the least-squares retrieval procedure of Sec.~3 and
Sec.~5A, with additional theoretical analysis in the Appendix.  We
placed an Schott BG18 colored glass filter (broad bandpass with peak transmission 
near $\sim 510$~nm) close to the source and orthogonal to 
the beam and took a single data
frame.  We then rotated the filter about the vertical axis by
approximately $30\degrees$ and took a second data frame.  Using these
single data frames in conjunction with the procedures described in
Sec.~3 and Sec.~5A, we derived polarization and intensity spectra,
shown in Fig.~10.  The average linear polarization for the region 500~nm
to 550~nm was 0.39\% and  6.2\% with 
standard deviations 0.02\% and 0.35\% respectively.  For a uniform
glass sheet of refractive index $n=1.5$ in air, we expect transmitted
light to be polarized 6.3\% for an incident angle of $30\degrees$.  We emphasize that these results were
obtained using a single data frame with no moving parts once the
calibration data had been acquired.

The third exercise was to demonstrate our ability to obtain full
Stokes polarimetry from a single data frame. For this, we used the
compound $3\degrees$ wedge pair used in the previous exercise,
followed by a compound $6\degrees$ wedge pair with fast axes at
$0\degrees$ and $90\degrees$. We removed the
quarter wave retarder.  As an interesting source of circularly
polarized light, we used a pair of plastic 3D cinema glasses.  These
glasses comprise a quarter wave sheet and polarizing sheet in
combination. Used as viewers, they either transmit or extinguish
circularly polarized light, depending on its sign. But, in reverse,
they produce circularly polarized light of opposite signs for each
``eye.''  We took single data frames through each eye in turn and
processed the data frames according to the methodology of Sec.~3 and
Sec.~5A, using the empirical suite of calibration data frames.  The results 
are shown in Fig.~11. The method produced excellent results on these
sources, yielding the expected extremely high level of polarization,
with opposite sign for the two eyes.  We also show the derived linear
polarization, to show that we are also measuring full the Stokes
vector in these single data frames.

We defer attempts to carry out precision polarimetry given the
rudimentary nature of our optical bench coupled to imperfect
calibration source availability. However, we believe that these
preliminary results are extremely encouraging and satisfy our desire
to provide a proof of concept in the laboratory and to demonstrate
that the desired polarimetric information can be retrieved from single
data frames.

\section{Conclusions}

We have described an approach to polarization measurement that uses no
moving parts and that relies on simple, robust optical
components. Either linear polarimetry or full Stokes polarimetry can
be carried out.  The method depends on the use of an area detector,
such as a CCD, with the light spread across a region of the detector.
If the system can be made photon-limited in sensitivity, this
spreading of the light improves the polarimetry, since typical CCD
well-depths are only of order $10^5$. With modest spreading of the
light, a single photon limited frame should be able to reach precision
of order $10^{-4}$ in polarization.  The influence of departures from
ideal circumstances still remains largely to be explored. Hence, we
do not know at this stage whether this approach will be able to
achieve extremely high accuracy. However the robustness and simplicity
of the components involved offers cause for optimism.  Other
approaches, such as the spectral modulation method for linear
polarimetry \cite{Snik2009}, offer alternative methods for static
polarimetry in hostile environments. Compared
to that approach, the methods presented here yield a cleaner
separation of the spectroscopy and polarimetry, at the expense of
additional detector surface area requirements. The methods may be applied in the UV or IR as well as in the visible wavelength range.

Since the entire polarization information is contained within a single
data frame, the method is well-suited to measuring the polarization of
transient sources and scenes where the polarimeter and target are in
rapid relative motion.  Since the optics are robust, simple and
require no moving parts, we anticipate that these methods will prove
useful for application in space.

\vspace{0.5in}
\noindent
{\bf Acknowledgments:} We acknowledge support from the STScI JWST
Director's Discretionary Research Fund JDF grant number
D0101.90152. STScI is operated by the Association for Universities for
Research in Astronomy, Inc., under NASA contract NAS5-26555. Patent
pending, all rights reserved.

\vspace{1.0in}

\noindent
{\bf Appendix: General Linear Least Squares Methods}

\vspace{12pt}

We follow Bevington\cite{Bevington1968,Bevington2002} and let
the general problem to be solved be
\begin{equation*}
y(x_i)=ai_c(x_i)+bq_c(x_i)+cu_c(x_i)+dv_c(x_i),
\end{equation*}
where measurements
$y_i$, either the intensity $I$ in the single beam case or
$(I_{\parallel}-I_{\perp})/(I_{\parallel}+I_{\perp})$ in the dual beam
case, are made at points $x_i$ and $y_i=y(x_i)+\epsilon_i$, with $y$
the true underlying value and $\epsilon_i$ its error (assumed random,
independent) at location $x_i$. The terms $i_c$, $q_c$, $u_c$, and
$v_c$ are trigonometric functions that encode the Stokes parameters
$I$, $Q$, $U$, and $V$ or $q$, $u$, and $v$, and their coefficients
$a$, $b$, $c$, and $d$ are the Stokes parameters to be derived. The
mapping and specific functions depend on the chosen configuration, but
all configurations discussed here can be expressed in this
way. Sometimes the functions are identically zero, implying no
sensitivity to that parameter.

The $\chi^2$ function is then
\begin{equation*}
\chi^2=\sum^{N}_{i=1}\epsilon_i^2/\sigma_i^2=\sum^{N}_{i=1}{1\over\sigma_i^2}[y_i-y(x_i)]^2=\sum^{N}_{i=1}{1\over\sigma_i^2}[y_i-ai_c(x_i)-bq_c(x_i)-cu_c(x_i)-dv_c(x_i)]^2,
\end{equation*}
and to solve, we set the partial derivatives of $\chi^2$ with respect to each of $a$, $b$, $c$, and $d$ equal to zero:
\begin{eqnarray*}
{\partial\chi^2\over\partial a}&=0=-2\sum {1\over\sigma_i^2} i_c\left(y_i-ai_c-bq_c-cu_c-dv_c\right),\\
{\partial\chi^2\over\partial b}&=0=-2\sum {1\over\sigma_i^2} q_c\left(y_i-ai_c-bq_c-cu_c-dv_c\right),\\
{\partial\chi^2\over\partial c}&=0=-2\sum {1\over\sigma_i^2} u_c\left(y_i-ai_c-bq_c-cu_c-dv_c\right),\\
{\partial\chi^2\over\partial d}&=0=-2\sum {1\over\sigma_i^2} v_c\left(y_i-ai_c-bq_c-cu_c-dv_c\right).\\
 \end{eqnarray*}
We require the curvature matrix $\mathbf{B}$ and summation vector $s_y$,
\begin{equation*}
{\bf B}\equiv\left( 
\begin{array}{cccc}
\sum{1\over\sigma_i^2}i_c^2   & \sum{1\over\sigma_i^2}i_cq_c   & \sum{1\over\sigma_i^2}i_cu_c    & \sum{1\over\sigma_i^2}i_cv_c   \\
\sum{1\over\sigma_i^2}i_cq_c & \sum{1\over\sigma_i^2}q_c^2    & \sum{1\over\sigma_i^2}q_cu_c  & \sum{1\over\sigma_i^2}q_cv_c  \\
\sum{1\over\sigma_i^2}i_cu_c & \sum{1\over\sigma_i^2}q_cu_c  & \sum{1\over\sigma_i^2}u_c^2    & \sum{1\over\sigma_i^2}u_cv_c  \\
\sum{1\over\sigma_i^2}i_cv_c & \sum{1\over\sigma_i^2}q_cv_c   & \sum{1\over\sigma_i^2}u_cv_c  & \sum{1\over\sigma_i^2}v_c^2    \\
\end{array}
 \right),
\end{equation*}
\begin{eqnarray*}
s_y\equiv\left(\sum{1\over\sigma_i^2}i_cy_i, \sum{1\over\sigma_i^2}q_cy_i, \sum{1\over\sigma_i^2}u_cy_i, \sum{1\over\sigma_i^2}v_cy_i\right),
\end{eqnarray*},
respectively. With this terminology, the least squares equations become
\begin{equation*}
s_y={\mathbf B}\cdot\left(
 \begin{array}{c}
 a\\
 b\\
 c\\
 d\\
 \end{array}
 \right).
\end{equation*}
We solve for the vector $(a,b,c,d)$,
\begin{equation*}
{\bf a}=(a,b,c,d)={\bf B}^{-1}\cdot s_y.
\end{equation*}
Following standard procedures,
e.g. \cite{Bevington1968},\cite{Bevington2002}, ignoring covariances,
the uncertainties on these parameters are
\begin{equation*}
\sigma^2_{a_i}=B_{ii}^{-1}
\end{equation*}
where $a_i$ represents $a$, $b$, $c$, or $d$, and $[B^{-1}]_{ii}$ is the corresponding diagonal term of ${\bf B}^{-1}$.

Our application is precision polarimetry, for which it is
presumed (\romannumeral1)~the degree of polarization is small, and
(\romannumeral2)~light levels are relatively high. Hence, the
intensity across the spatial segment is approximately constant and
obeys Poisson counting statistics. That is, we assume the uncertainty
is the same for each bin, $\sigma_i = \sigma =(N_{\gamma}/nx)^{(1/2)}$
where $N_{\gamma}$ is the total number of detected photons, and $nx$ is
the number of bins across which the photons are distributed, i.e., the number of
$x$ sampling points. For cases where the trigonometric functions that are embodied by $i_c$,
$q_c$, $u_c$, $v_c$ are orthogonal (we approximate the summations by
integrals over integer numbers of periods), ${\bf B}$ is diagonal. 
Hence its inverse is also diagonal, and the Stokes parameter solutions are
independent of one another.

We choose as a simple example the single wedge with its fast axis at
45$\degrees$ to the slit direction, which, in turn, defines the
direction for Stokes $Q$. In general, we use Mueller matrix algebra to
solve for the system. As in Sec. 2.A above, Eq.~(1) gives the
expression for the intensity at points $x_i$: $y_i\equiv
I(x_i)=0.5(I-Q\cos\phi_1-U\sin\sin\phi_i)$ where $\phi_i=2\pi
(x_i/X)$. In the formalism above,
$y_i=ai_c(x_i)+bq_c(x_i)+cu_c(x_i)+dv_c(x_i)$ so $i_c(x_i)=0.5$,
$q_c(x_i)=0.5\cos(2\pi x_i/X)$, $u_c(x_i)=0.5\sin(2\pi x_i/X)$ and
$v_c(x_i)=0$.  In the absence of $V$ the matrix ${\bf B}$ is reduced
to the $3\times 3$ matrix
\begin{equation*}
{\bf B}\equiv{1\over\sigma^2}\left( 
\begin{array}{ccc}
\sum i_c^2   & \sum i_cq_c   & \sum i_cu_c     \\
\sum i_cq_c & \sum q_c^2    & \sum q_cu_c   \\
\sum i_cu_c & \sum q_cu_c  & \sum u_c^2     \\
\end{array}
 \right).
\end{equation*}
The summations run across $nx$ pixels.  Applying the expressions for the $q_c$, $u_c$, and $v_c$, we have
\begin{equation*}
{\bf B}={1\over 4\sigma^2}\left( 
\begin{array}{ccc}
nx   & \sum \cos(2\pi x_i/X)   & \sum \sin(2\pi x_i/X)     \\
\sum \cos(2\pi x_i/X) & \sum \cos^2(2\pi x_i/X)    & \sum \cos(2\pi x_i/X)\sin(2\pi x_i/X)   \\
\sum \sin(2\pi x_i/X) & \sum \cos(2\pi x_i/X)\sin(2\pi x_i/X)  & \sum \sin^2(2\pi x_i/X)     \\
\end{array}
 \right).
\end{equation*}
We assume the summations cover an integer number of periods and approximate the sums using integrals, using
$\int_0^Xf(x)dx\approx \Delta x\sum_1^{nx} f(x_i)$, where $\Delta x = X/nx$ (the width of a bin in $x$). Hence, $\sum f(x_i)\approx {nx\over X}\int f(x)dx$.
Thus, it can be shown for this example that
\begin{equation*}
{\bf B}={nx\over 8\sigma^2}\left( 
\begin{array}{ccc}
2  & 0  & 0   \\
0 & 1   & 0  \\
0 & 0  & 1   \\
\end{array}
 \right),
\end{equation*}
and
\begin{equation*}
{\bf B}^{-1}={4\sigma^2\over nx}\left( 
\begin{array}{ccc}
1  & 0  & 0   \\
0 & 2   & 0  \\
0 & 0  & 2   \\
\end{array}
 \right).
\end{equation*}
 Now, we can go back to the expression for $s_y$ and solve for the
 Stokes parameters, noting that $\sum y_i = N_{tot}$, the total number
 of photons collected, and $s_y = {1\over 2\sigma^2} \left[\sum y_i, \sum
 \cos(2\pi x_i/X)y_i, \sum \sin(2\pi x_i/X)y_i \right]$, omitting the
 zero $V$ term.  Hence the solutions from $(a,b,c,d)={\bf B}^{-1}\cdot s_y$
 are
\begin{eqnarray*}
I\equiv a &= \left({N_{tot}\over 2\sigma^2}\right)\left({4\sigma^2\over nx}\right)={2N_{tot}\over nx}=2\langle y\rangle,\\
Q\equiv b &= \left({4\over nx}\right)\sum y_i \cos(2\pi x_i/X),\\
U\equiv c &= \left({4\over nx}\right)\sum y_i \sin(2\pi x_i/X).
\end{eqnarray*}
 Similarly, we can derive the uncertainties of the Stokes
 parameters. The uncertainty $\sigma$ is given by
 $\sigma^2=N_{tot}/nx$ and is assumed to be constant. Hence, reading directly from
 the expression for ${\bf B}^{-1}$, we have
\begin{eqnarray*}
\sigma(I)\equiv\sigma_a &={2\sqrt{N_{tot}}\over nx},\\
\sigma(Q)\equiv\sigma_b &={2\sqrt{2N_{tot}}\over nx},\\
\sigma(U)\equiv\sigma_c &={2\sqrt{2N_{tot}}\over nx}.
\end{eqnarray*}
 The uncertainties in the normalized Stokes parameters $q$ and $u$ are 
\begin{equation*}
\sigma(q)=\sigma(u)=\sqrt{{2\over N_{tot}}}.
\end{equation*}
Ignoring bias terms, it follows that the uncertainty on the degree of polarization is
\begin{equation*}
\sigma(p)=\sqrt{{2\over N_{tot}}}.
\end{equation*}
 
The expressions for the trigonometric functions $i_c$, $q_c$, $u_c$,
$v_c$ depend on the configuration and whether a dual beam formalism is
adopted or not. We derived the expressions for these functions using
Mueller matrix algebra for a selection of configurations, as presented
in Table~1. In a similar fashion to this example, though with more
complex manipulations, we can analytically invert the corresponding
curvature matrix to derive both the solution and the uncertainty estimates,
taking only the diagonal terms as the uncertainty. There are cases
where the off-diagonal terms of ${\bf B}$ are non-zero, as discussed
in the text.

\begin{table}
\caption{Coefficients of Stokes parameters for different wedge configurations}
\begin{center}
{\tiny \begin{tabular}{llrrrr}
\hline
Wedges$^{\hbox{a}}$ & Beam & $i_c$ & $q_c$ & $u_c$ & $v_c$ \\
\hline
$qw$ & single  &0.5  &  $0.5\cos\phi$ & $0.5\sin\phi$ &  \\
 &dual &    &  $\cos\phi$ & $\sin\phi$ &   \\
$qww^\prime$ & single &  0.5 & $0.5\cos 2\phi$ & $0.5\sin 2\phi$ &   \\
 &dual&   & $\cos 2\phi$ & $\sin 2\phi$ &   \\
$wW$ & single&  0.5  &  $0.5(\cos\phi\cos 2\theta+\sin\phi\sin 2\phi\sin 2\theta)$ & $0.5\cos 2\phi\sin 2\theta$ & $0.5(\cos\phi\sin 2\phi\sin 2\theta-\sin\phi\cos 2\theta)$ \\
 & dual &   &  $\cos\phi\cos 2\theta+\sin\phi\sin 2\phi\sin 2\theta$ & $\cos 2\phi\sin 2\theta$ & $\cos\phi\sin 2\phi\sin 2\theta-\sin\phi\cos 2\theta$ \\
$wW^\prime$ & single &  0.5  &  $0.5(\cos\phi\cos 2\theta+\sin\phi\sin(\zeta -2\phi)\sin 2\theta)$ & $0.5\cos(\zeta-2\phi)\sin 2\theta$ & $0.5(\cos\phi\sin(\zeta -2\phi)\sin 2\theta-\sin\phi\cos 2\theta)$ \\
 & dual&   &  $\cos\phi\cos 2\theta+\sin\phi\sin(\zeta -2\phi)\sin 2\theta$ & $\cos(\zeta-2\phi)\sin 2\theta$ & $\cos\phi\sin(\zeta -2\phi)\sin 2\theta-\sin\phi\cos 2\theta$ \\
$ww^\prime WW^\prime$ & single & 0.5  &   $0.5(\cos 2\phi\cos 2\theta+\sin 2\phi\sin 4\phi\sin 2\theta)$ & $0.5\cos 4\phi\sin 2\theta$ & $0.5(\cos 2\phi\sin 4\phi\sin 2\theta-\sin 2\phi\cos 2\theta)$ \\
              & dual&  &   $\cos 2\phi\cos 2\theta+\sin 2\phi\sin 4\phi\sin 2\theta$ & $cos 4\phi\sin 2\theta$ & $\cos 2\phi\sin 4\phi\sin 2\theta-\sin 2\phi\cos 2\theta$ \\

\end{tabular}}
\end{center}
\footnotesize{ $^a$Notation for wedge configurations: $q$ denotes
  optional quarter wave retarder, $w$ denotes thickness gradient
  $\phi_w= 2\pi x/X$ and $W$ denotes twice the thickness gradient
  $\phi_W= 4\pi x/X$. A primed symbol denotes antiparallel wedge
  direction relative to unprimed. Wedges $w$ have fast axis at
  $45\degrees$ to the slit, $w^\prime$ at $-45\degrees$; wedges $W$
  have fast axis at $0\degrees$, except in combination $WW^\prime$, when they
  are at $0\circ$ and $90\degrees$ respectively.}\normalsize
\end{table}

\begin{table}
\caption{Error estimates for normalized Stokes parameters for different wedge configurations}
\begin{center}
{\tiny \begin{tabular}{llccc}
\hline
Wedges&Beam &  $\sigma(q)$ & $\sigma(u)$ & $\sigma(v)$ \\
\hline
$qw$ & single & $(2/ N_{tot})^{1/2}$ & $(2/ N_{tot})^{1/2}$  &  \\
&dual &$(2/ N_{tot})^{1/2}$  &  $(2/ N_{tot})^{1/2}$ \\
$qww^\prime$ &single & $(2/ N_{tot})^{1/2}$  &$(2/ N_{tot})^{1/2}$   &  \\
&dual &$(2/ N_{tot})^{1/2}$  &  $(2/ N_{tot})^{1/2}$ \\
$wW$&single & $2(2/N_{tot})^{1/2}(3+\cos 4\theta +2\sin 4\theta)^{-1/2}$  &$(2/N_{tot})^{1/2}/|\sin 2\theta|$ & $2(2/N_{tot})^{1/2}(3+\cos 4\theta -2\sin 4\theta)^{-1/2}$   \\
&dual&  $2(2/N{tot})^{1/2}(3+\cos 4\theta +2\sin 4\theta)^{-1/2}$   & $(2/N_{tot})^{1/2}/|\sin 2\theta|$& $2(2/N{tot})^{1/2}(3+\cos 4\theta -2\sin 4\theta)^{-1/2}$    \\
$wW^\prime$&single & $4/N_{tot}^{1/2}\left({3+\cos 4\theta + 2\cos\zeta\sin 4\theta\over 15+12\cos 4\theta + 5\cos 8\theta}\right)^{1/2}$ & $(2/N_{tot})^{1/2}/|\sin 2\theta|$& $4/N_{tot}^{1/2}\left({3+\cos 4\theta - 2\cos\zeta\sin 4\theta\over 15+12\cos 4\theta + 5\cos 8\theta}\right)^{1/2}$\\
&dual &  $4/N_{tot}^{1/2}\left({3+\cos 4\theta + 2\cos\zeta\sin 4\theta\over 15+12\cos 4\theta + 5\cos 8\theta}\right)^{1/2}$  &$(2/N_{tot})^{1/2}/|\sin 2\theta|$ & $4/N_{tot}^{1/2}\left({3+\cos 4\theta - 2\cos\zeta\sin 4\theta\over 15+12\cos 4\theta + 5\cos 8\theta}\right)^{1/2}$ \\
$ww^\prime WW^\prime$&single&  $2(2/N{tot})^{1/2}(3+\cos 4\theta +2\sin 4\theta)^{-1/2}$   & $(2/N_{tot})^{1/2}/|\sin 2\theta|$& $2(2/N{tot})^{1/2}(3+\cos 4\theta -2\sin 4\theta)^{-1/2}$     \\
&dual&  $2(2/N{tot})^{1/2}(3+\cos 4\theta +2\sin 4\theta)^{-1/2}$   & $(2/N_{tot})^{1/2}/|\sin 2\theta|$& $2(2/N{tot})^{1/2}(3+\cos 4\theta -2\sin 4\theta)^{-1/2}$   \\

\end{tabular}}
\end{center}
\end{table}

\begin{table}
\caption{Error estimates for unnormalized Stokes parameters for different wedge configurations}
\begin{center}
{\tiny \begin{tabular}{llccccccc}
\hline
Weges&Beam & $\sigma(I)$ & $\sigma(Q)$ & $\sigma(U)$ & $\sigma(V)$  \\
\hline
$qw$ & single & ${2N_{tot}^{1/2}/ nx}$ & $2(2N_{tot})^{1/2}/ nx$ & $2(2N_{tot})^{1/2}/ nx$ &  \\
&dual && & &  \\
$qww^\prime$ &single & ${2N_{tot}^{1/2}/ nx}$  &$2(2N_{tot})^{1/2}/ nx$ & $2(2N_{tot})^{1/2}/ nx$&  \\
&dual&    & & &   \\
$wW$&single & $2N_{tot}^{1/2}/ nx$  &${4(2N_{tot}/(3+\cos 4\theta +2\sin 4\theta))^{1/2}/ nx}$  &${2(2N_{tot})^{1/2}/(nx|\sin 2\theta|)}$ & ${4(2N_{tot}/(3+\cos 4\theta -2\sin 4\theta))^{1/2}/ nx}$  \\
&dual&  & & &    \\
$wW^\prime$&single & $2N_{tot}^{1/2}/ nx$& ${8\over nx}\left(N_{tot}{3+\cos 4\theta + 2\cos\zeta\sin 4\theta\over 15+12\cos 4\theta + 5\cos 8\theta}\right)^{1/2}$ &$2(2N_{tot})^{1/2}/(nx|\sin 2\theta|)$  &${8\over nx}\left(N_{tot}{3+\cos 4\theta - 2\cos\zeta\sin 4\theta\over 15+12\cos 4\theta + 5\cos 8\theta}\right)^{1/2}$\\
&dual &   & & & & &  \\
$ww^\prime WW^\prime$&single&  $2N_{tot}^{1/2}/ nx$& ${4(2N_{tot}/(3+\cos 4\theta +2\sin 4\theta))^{1/2}/ nx}$  &${2(2N_{tot})^{1/2}/(nx|\sin 2\theta|)}$& ${4(2N_{tot}/(3+\cos 4\theta -2\sin 4\theta))^{1/2}/ nx}$   \\
&dual&  & & &    \\

\end{tabular}}
\end{center}
\end{table}

\clearpage

\bibliographystyle{osajnl.bst}

\clearpage


  \begin{figure}[htbp]
  \centering
  \includegraphics[width=14cm]{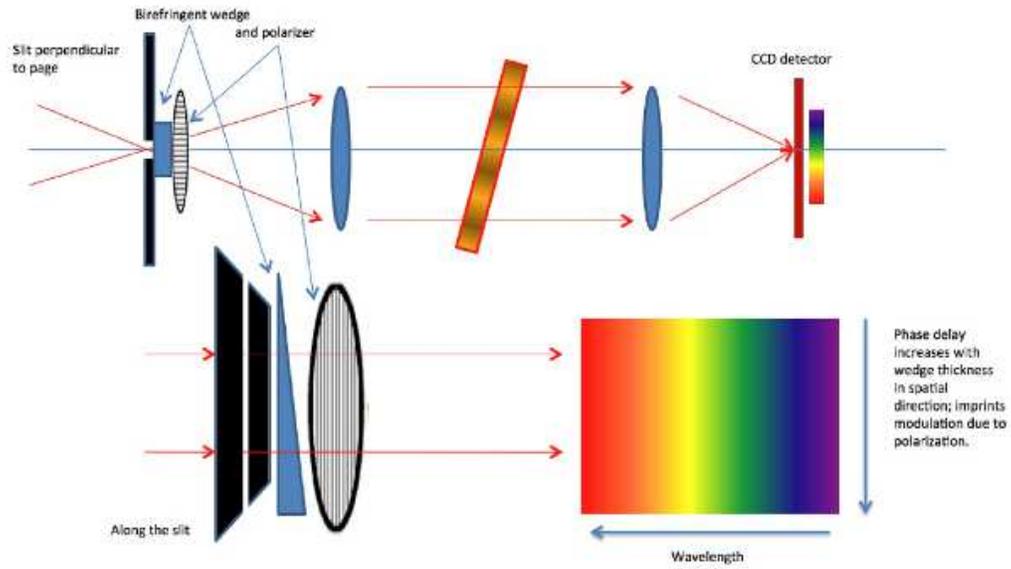}
   \caption{ Illustration of the optical bench layout. Light enters a
     spatially elongated slit, passes through a birefringent wedge or
     wedges and polarization analyzer, and then enters a conventional
     long-slit spectrograph. A quarter wave retarder (not shown) may
     be inserted before the wedge.}
  \end{figure}

 \begin{figure}[htbp]
  \centering
  \includegraphics[width=11cm]{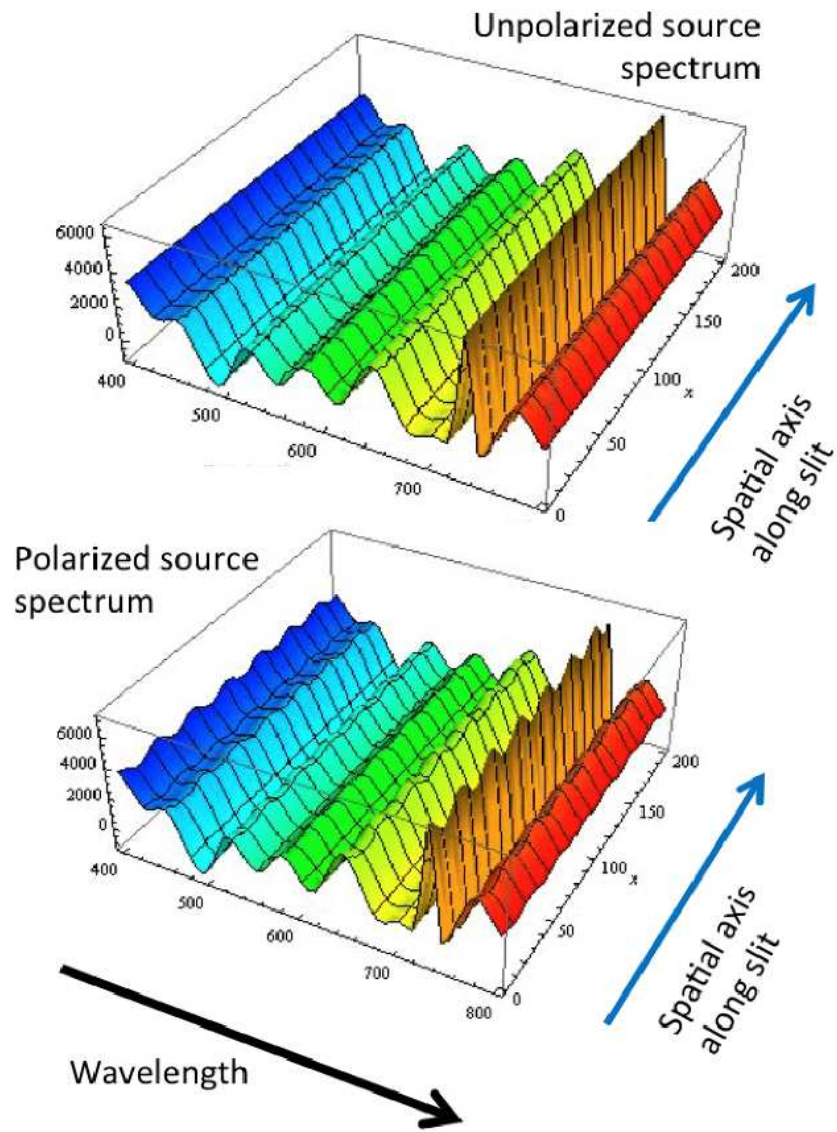}
   \caption{ Concept --- the polarization optics imprint an amplitude
     modulation on the dimension orthogonal to the dispersion
     direction of the spectrograph. Typically, this direction corresponds to the
     spatial dimension along the slit.}
  \end{figure}

 \begin{figure}[htbp]
  \centering
  \includegraphics[width=9cm]{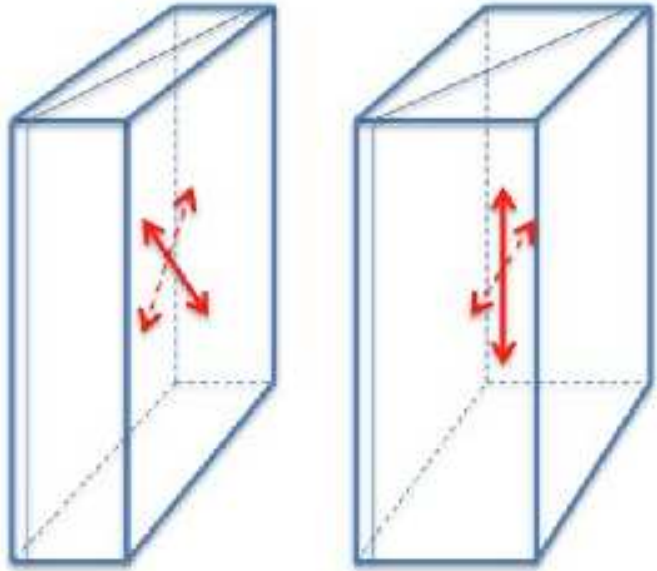}
   \caption{ Illustration of the compound birefringent wedges. A single
     compound optic may be used for linear polarimetry (left) and
     double (both) for full Stokes. The fast axes run at $\pm
     45^{\circ}$ in the left optic, horizontal and vertical in the
     right optic, and the slit direction is horizontal or
     vertical.}
  \end{figure}

\begin{figure}[htbp]
  \centering
  \includegraphics[width=10cm]{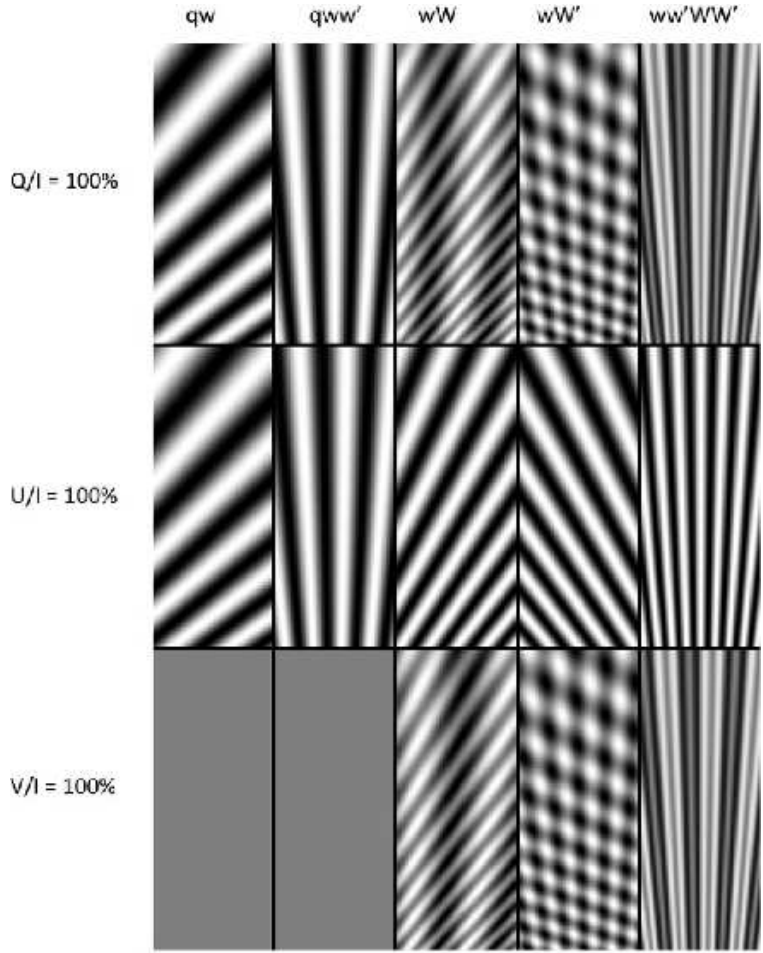}
   \caption{ Example theoretical data frames for each of the
     configurations discussed in the text when viewed with
     100\% polarized light. Each panel has $x$ running horizontally
     and wavelength vertically, increasing up. Parameters correspond
     to 2~mm in $x$ of a $3\degrees$ quartz wedge set,
     running from 450~nm to 750~nm. Top row shows 100\% Stokes $Q$,
     middle row 100\% Stokes $U$ and bottom row 100\% Stokes
     $V$. Left to right, in the notation of Tables~1--3, the
     configurations are $qw$, $qww^\prime$, $wW$, $wW^\prime$, and
     $ww^\prime WW^\prime$. Note that if the quarter wave retarder were
     omitted in the first two columns, $U$ and $V$, which show no
     sensitivity with the quarter wave retarder, would be
     interchanged. For the first two columns, the fast axis is set at
     $0\degrees$ and for the remaining three at $45\degrees$ (see
     text).}
  \end{figure}

\begin{figure}[htbp]
  \centering
  \includegraphics[width=11cm]{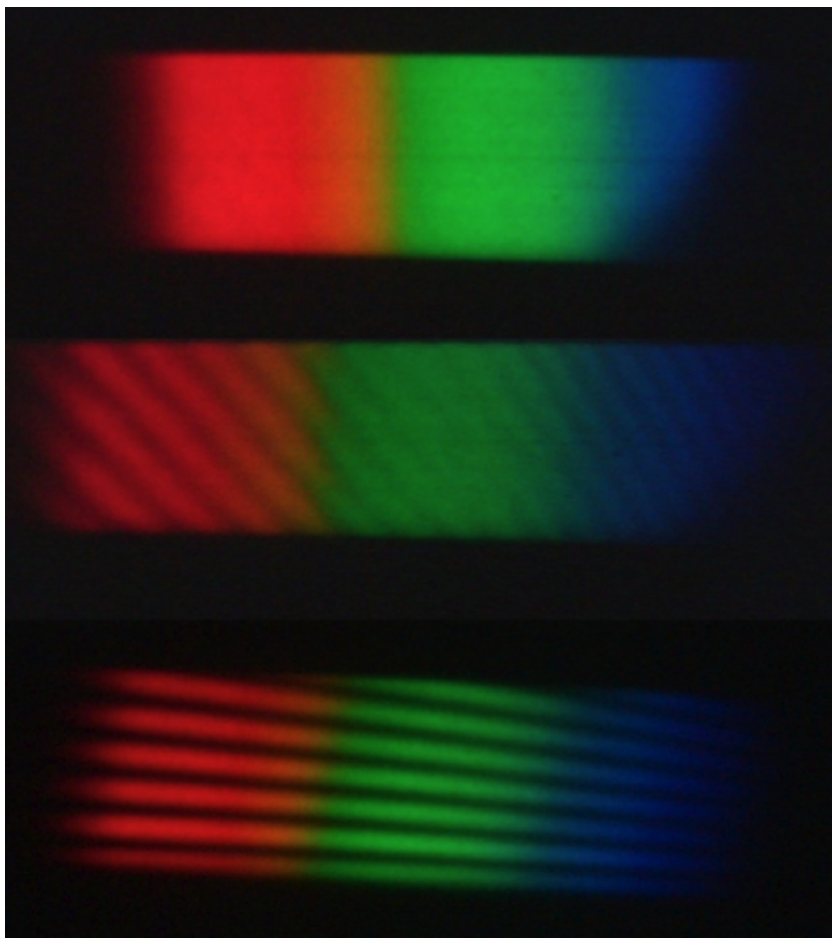}
   \caption{ Actual spectra obtained with a preliminary test optical
     bench: (upper) a spectrum obtained with unpolarized light, (center)
     a spectrum with one quartz birefringent wedge and 100\% linearly
     polarized light, and (lower) a spectrum obtained with two quartz
     birefringent wedges reversed as in the manner of the compound
     optics with 100\% linearly polarized light.}
  \end{figure}

\begin{figure}[htbp]
  \centering
  \includegraphics[width=11cm]{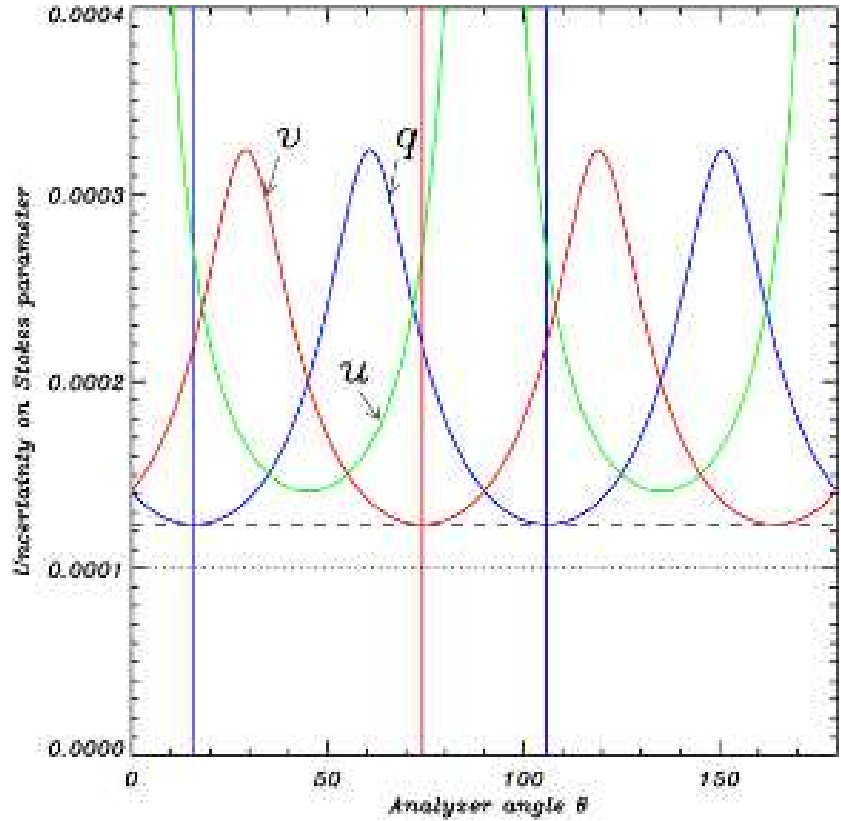}
   \caption{ The uncertainties on normalized Stokes
     parameters for double wedge pair configuration, Sec.~4.B.3. Blue
     is Stokes $q$, green is $u$, and red is $v$. The vertical lines
     indicate the positions of the minima for $q$ and $v$, the
     horizontal lines indicates $1/\sqrt{N_{tot}}$ (dotted), and the
     analytically-determined minimum $1.24\times$ higher (dashed).}
  \end{figure}

\begin{figure}[htbp]
  \centering
  \includegraphics[width=11cm]{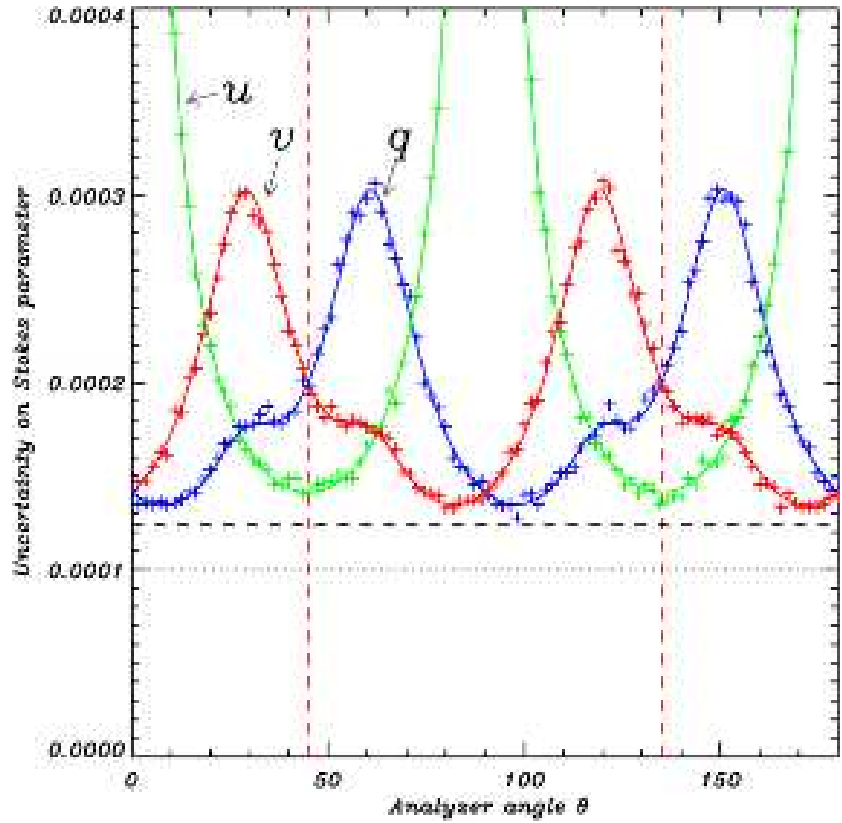}
   \caption{ The uncertainties on normalized Stokes
     parameters for double wedge pair configuration, Sec.~4.B.3 with
     miscentering of order $1/32\times$ the spatial distance of one
     wavelength of retardance. Blue is Stokes $q$, green is $u$, and
     red is $v$. The vertical lines indicate the positions of the
     analyzer angles that have no formal covariance, which is
     independent of the miscentering. Smooth lines through the
     simulated data are the analytic solutions ignoring covariance
     terms, while the plus signs are the results of Monte-Carlo
     simulations. The horizontal lines are as in the previous figure.}
  \end{figure}

\begin{figure}[htbp]
  \centering
  \includegraphics[width=6.5in]{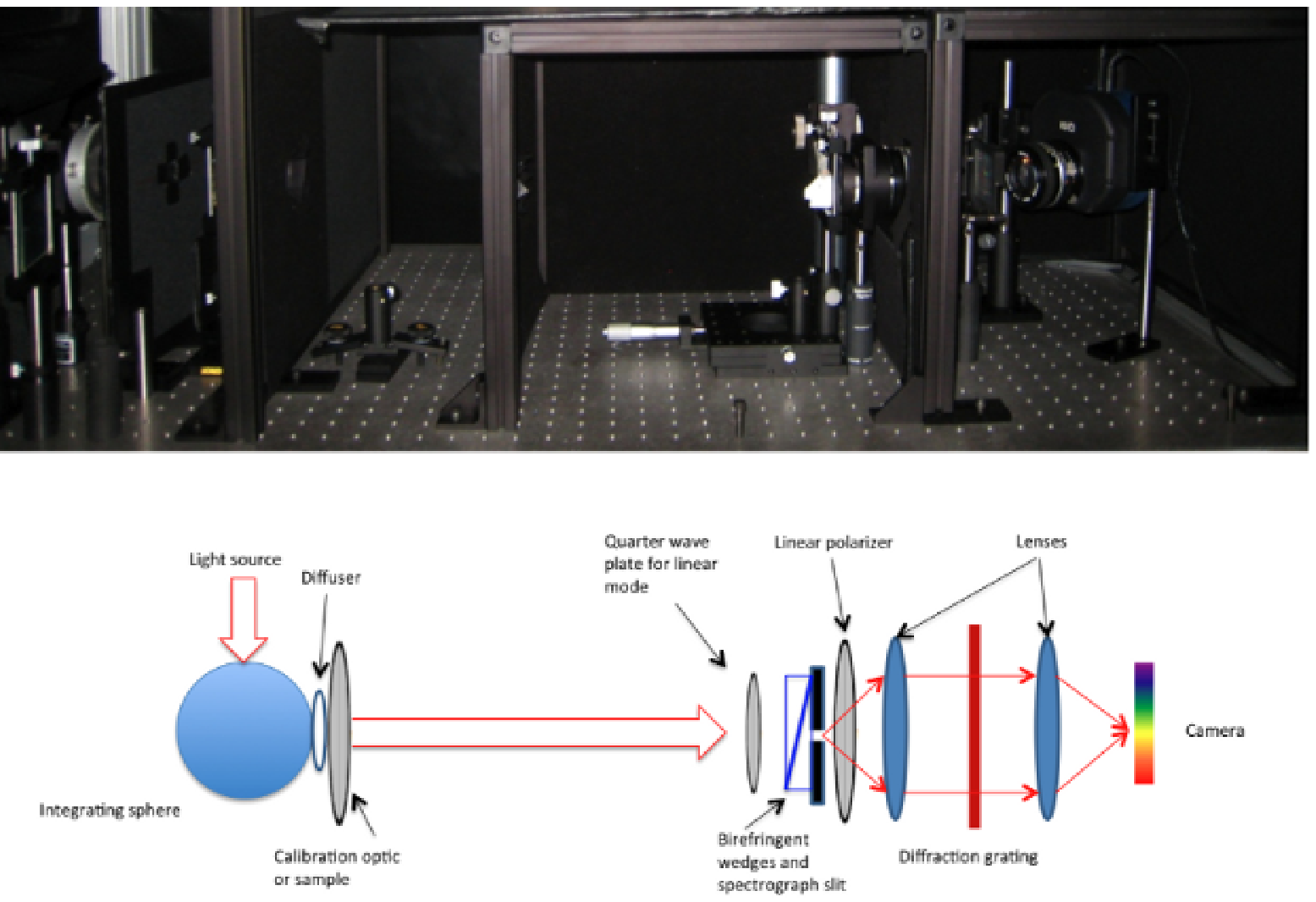}
   \caption{Laboratory optical bench layout as implemented. Light
     enters an integrating sphere, and illuminates a diffusing screen
     on exit. It then passes through calibration or sample components,
     before entering the long slit spectrograph with its associated
     polarization components as described in the text. The upper panel
     shows the actual optical bench with baffles and boxes removed for
     visibility. The integrating sphere is to the left, and the camera
     to the right.}
  \end{figure}

\begin{figure}[htbp]
  \centering
  \includegraphics[width=10cm]{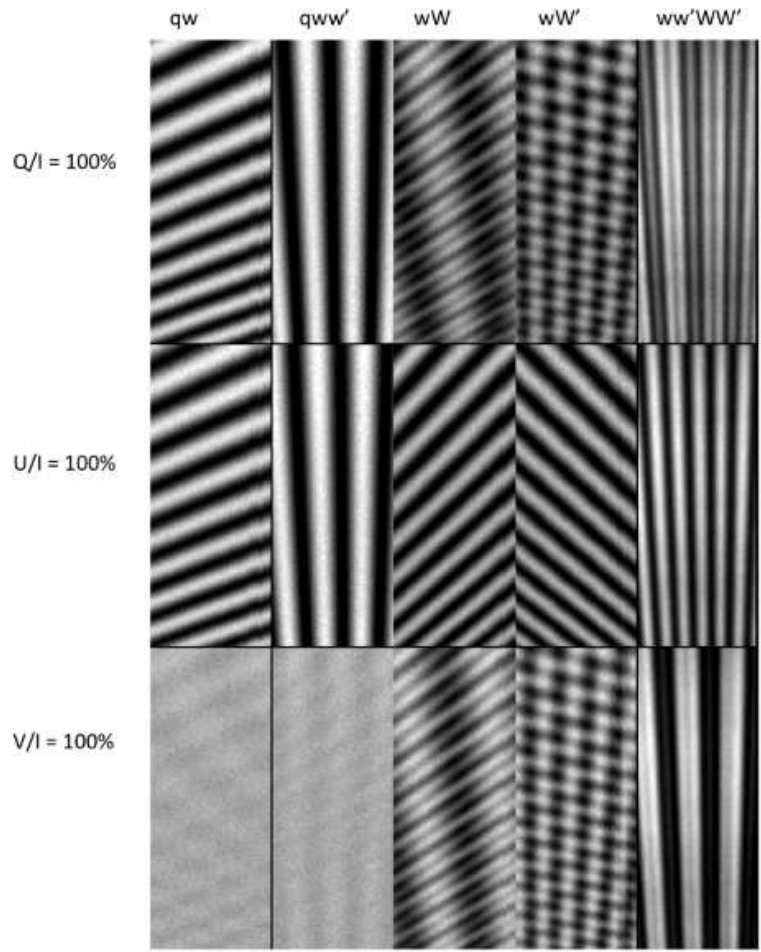}
   \caption{Example data frames for each of the configurations
     discussed in the text when viewed with $\approx
     100$\% polarized light, obtained in the laboratory. The rows and columns 
     correspond to those shown in Fig.~4. }
  \end{figure}

\begin{figure}[htbp]
  \centering
  \includegraphics[width=12cm]{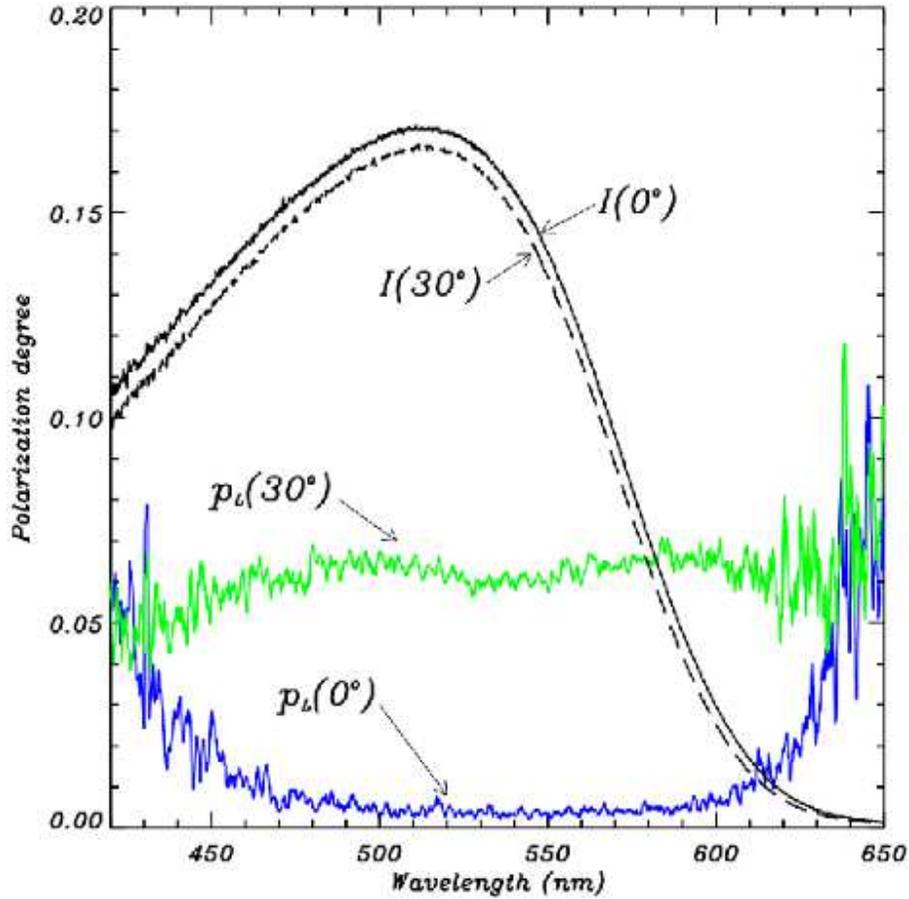}
   \caption{Retrieved polarization curves for a BG18 colored glass
     filter presented orthogonal to the beam, blue, and at an angle
     tilted by $\approx 30\degrees$ to orthogonal, green, observed
     using the $wW^\prime$ configuration. At right angles, we expect no
     polarization, and inclined at $30\degrees$, approximately $6.3$\%,
     consistent with the least squares retrieval. The black curves
     show arbitrarily normalized throughputs for the two
     configurations (solid, orthogonal and dotted, inclined) derived
     from the data, serving to illustrate that we also obtain full
     Stokes~$I$ spectroscopy using these methods.}
  \end{figure}

\begin{figure}[htbp]
  \centering
     \includegraphics[width=12cm]{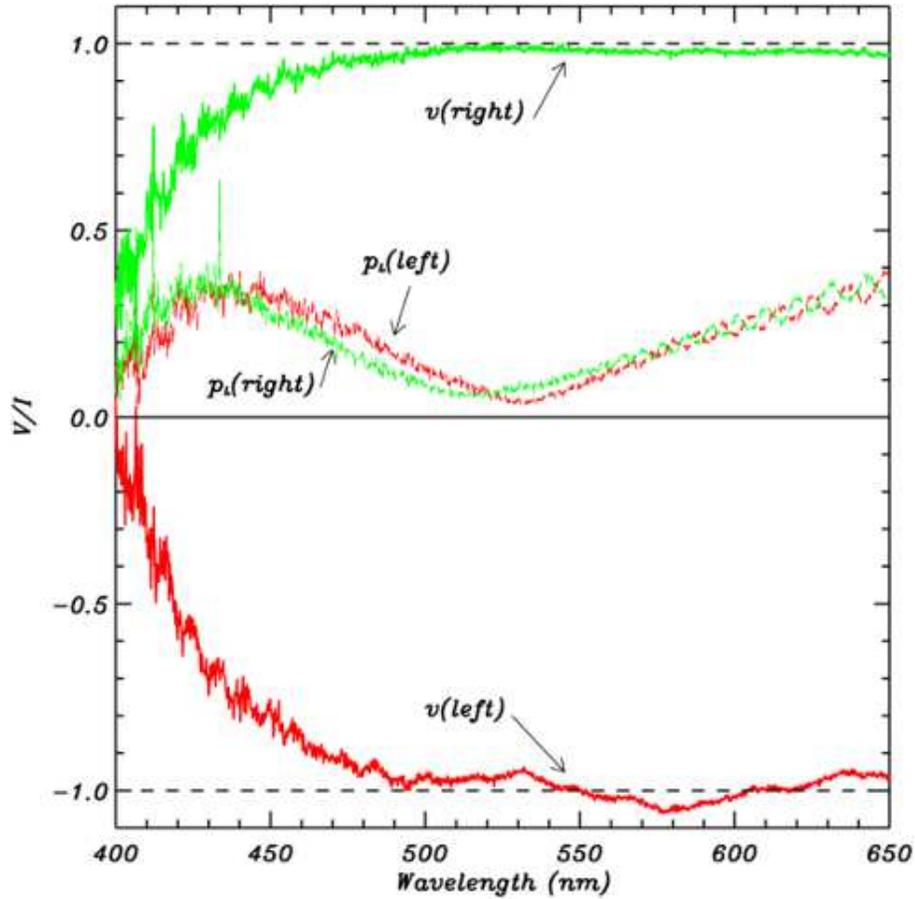}

   \caption{Retrieved circular polarization for a pair of polarizing
     cinema 3D glasses, expected to exhibit 100\% Stokes $V$ left
     and right circularly polarized light for the left and right eyes,
     measured using the $ww^\prime WW^\prime$ configuration. The
     retrieval is consistent with expectations. For completeness, and
     to illustrate that we obtain full Stokes polarimetry from a
     single data frame, the dashed lines show the retrieved degree of
     linear polarization.}
  \end{figure}

\end{document}